\documentclass[fleqn,usenatbib]{mnras}

\usepackage{newtxtext,newtxmath}

\usepackage[T1]{fontenc}

\DeclareRobustCommand{\VAN}[3]{#2}
\let\VANthebibliography\thebibliography
\def\thebibliography{\DeclareRobustCommand{\VAN}[3]{##3}\VANthebibliography}

\usepackage{graphicx}	
\usepackage{amsmath}	
\usepackage{color}
\usepackage{comment}
\usepackage{gensymb}
\usepackage{natbib}
\usepackage{caption}

\title[The flickering radio jet of A0620-00]{The flickering radio jet from the quiescent black hole X-ray binary A0620-00}

\author[D.\ L.\ dePolo et al.]{Donna L.\ dePolo,$^{1}$\thanks{E-mail: donna.depolo@nevada.unr.edu}
Richard M.\ Plotkin,$^{1}$
James C.\ A.\ Miller-Jones,$^{2}$
Jay Strader,$^{3}$ 
Thomas J.\ Maccarone,$^{4}$ \newauthor
Tyrone N.\ O'Doherty,$^{2}$
Laura Chomiuk,$^{3}$
Elena Gallo$^{5}$
\\
$^{1}$Physics Department, University of Nevada, Reno, 1664 N. Virginia St., Reno, NV 89557, USA\\
$^{2}$International Centre for Radio Astronomy Research-Curtin University, Perth, WA 6845, Australia\\
$^{3}$Center for Data Intensive and Time Domain Astronomy, Department of Physics and Astronomy, Michigan State University, East Lansing, MI 48824, USA\\
$^{4}$Department of Physics and Astronomy, Texas Tech University, Lubbock TX 79409-1051, USA\\
$^{5}$Department of Astronomy, University of Michigan, 1085 S University, Ann Arbor, MI 48109, USA
}

\date{Accepted XXX. Received YYY; in original form ZZZ}

\pubyear{2022}

\begin{document}
\label{firstpage}
\pagerange{\pageref{firstpage}--\pageref{lastpage}}
\maketitle

\begin{abstract}
Weakly accreting black hole X-ray binaries launch compact radio jets that persist even in the quiescent spectral state, at X-ray luminosities $\lesssim 10^{-5}$ of the Eddington luminosity. However, radio continuum emission has been detected from only a few of these quiescent systems, and little is known about  their radio variability. Jet variability can lead to misclassification of accreting compact objects in quiescence, and affects the detectability of black hole X-ray binaries in next-generation radio surveys. Here we present the results of a radio monitoring campaign of A0620$-$00, one of the best-studied and least-luminous known quiescent black hole X-ray binaries. We observed A0620$-$00 at 9.8 GHz using the Karl G.\ Jansky Very Large Array on 31 epochs from 2017 to 2020, detecting the source $\sim 75\%$ of the time. We see significant variability over all timescales sampled, and the observed flux densities follow a lognormal distribution with $\mu=12.5\ \mu$Jy and $\sigma = 0.22$ dex. In no epoch was A0620$-$00 as bright as in 2005 ($51\pm7 \mu$Jy), implying either that this original detection was obtained during an unusually bright flare, or that
the system is fading in the radio over time.
We present tentative evidence that the quiescent radio emission from A0620$-$00 is less variable than that of V404 Cyg, the only other black hole binary with comparable data. Given that V404 Cyg has a jet radio luminosity $\sim 20$ times  higher than A0620$-$00, this comparison could suggest that less luminous jets are less variable in quiescence.

\end{abstract}




\section{Introduction}

    Transient black hole X-ray binaries (BHXBs) spend the majority of their time experiencing very low levels of accretion in a state of quiescence, which we define by X-ray luminosities $L_X \lesssim 10^{-5}~L_{\rm Edd}$\footnote{$L_{\rm Edd}=1.3\times10^{38} \left(M_{\rm BH}/M_\odot\right)$ is the Eddington luminosity.} 
    \citep{2013PlotkinGalloJonker}. There is growing evidence that quiescent BHXBs launch relativistic outflows in the form of partially self-absorbed synchrotron jets  \citep{blandford79}, analogous to the compact radio jets ubiquitously observed from outbursting BHXBs in the hard X-ray spectral state \citep{fender01, corbel02}.     However, due to their low radio luminosities, we currently have radio detections of jets from only seven quiescent systems: A0620$-$00 \citep{2006GalloFenderMillerJones, 2018DincerBailyn}, BW Cir \citep{2021PlotkinBahramian}, GX 339-4 \citep{tremou20}, MAXI J1348$-$630 \citep{carotenuto22},  MWC 656 \citep{dzib15, ribo17}, V404 Cygni \citep{hjellming00, gallo05, hynes09, rana16}, and  XTE J1118+480 \citep{gallo14}, which launch  jets with  radio luminosities ranging from $\approx$$10^{26}-10^{28}$ erg  s$^{-1}$. Lack of  radio detections from other quiescent systems does not prove the absence of jets, however, since it is plausible that current radio observations are simply not sensitive enough \citep[e.g.,][]{miller-jones11, plotkin16, rodriguez20}. 
    One factor, among several, that makes detecting quiescent radio jets challenging is that their level of flux variability is poorly constrained. In turn, interpreting radio observations typically requires coordinating  radio campaigns with (strictly simultaneous) observations at other wavebands, especially in the X-ray \citep[see, e.g.,][and references therein]{rodriguez20}.
    
   The vast majority of current BHXBs were identified via enhanced X-ray emission during an outburst, which leads to biases in the current BHXB population to objects with, e.g.,  longer orbital periods, lower inclinations, lower black hole masses,  and/or lower line-of-sight absorption \citep[e.g.,][]{narayan05, jonker21}.  Identifying BHXBs in quiescence can alleviate some biases,  and various strategies have been suggested (and executed) to find quiescent BHXBs via electromagnetic signatures of accretion \citep[e.g.,][]{2005Maccarone, jonker14, casares18}, via dynamical searches of detached systems \citep{giesers19, thompson19}, and (for binary black holes) through gravitational wave events \citep{abbott16}.
   Radio surveys to discover quiescent BHXBs via their relativistic jets have been suggested and performed  \citep[e.g.,][]{2005Maccarone, shishkovsky20, tudor22}, and such  surveys are revealing intriguing radio-selected candidates \citep[e.g.,][]{tetarenko16bhxb}, particularly within globular clusters \citep[e.g.,][]{strader12, chomiuk13, miller-jones15, shishkovsky18, urquhart20, zhao20}.  However, it is practically impossible to coordinate the requisite strictly simultaneous multiwavelength observations with the initial radio survey observations. Preliminary identifications of radio-selected BHXB candidates can become more reliable by improving characterisation of the level of variability displayed by quiescent radio jets.

    \citet{2019PlotkinMillerJones} performed the first large radio variability study of a quiescent BHXB, focusing on V404 Cygni, which is one of the brightest BHXBs in quiescence (X-ray luminosity $L_{\rm X} \approx 10^{33}$ erg s$^{-1}$, or equivalently $\approx 10^{-6} L_{\rm Edd}$; \citealt{2014BernardiniCackett}). From analyzing 24 years worth of  observations in the Karl G. Jansky Very Large Array (VLA) archive, they found that flux densities of V404 Cygni follow a lognormal distribution with a mean $\left<\log \left(f_\nu/{\rm  mJy}\right)\right>=-0.53$ and a standard deviation of $\pm$0.30 dex at 8.4 GHz (based on 89 observations), corresponding to an average radio luminosity $L_{\rm R} \approx 10^{28}$ erg s$^{-1}$. They also concluded that factor of 2-4 variations are common on timescales ranging from minutes to decades, and those variations  are consistent with either flicker noise or white noise.  Combining these results with previously identified flaring behaviour on timescales of tens of minutes to hours \citep[e.g.,][]{miller-jones08} suggests that the radio variability of V404 Cygni is consistent with a damped random walk, which may be caused by shocks propagating through a steady, compact jet \citep{2019PlotkinMillerJones}.
    
    In an effort to continue the characterisation of quiescent radio jet variations, we repeat a similar analysis on the BHXB A0620$-$00 in quiescence. A0620$-$00 is a natural choice for the next target in this type of analysis because it is $\approx$100-200 times less luminous in the X-ray compared to V404 Cygni, such that A0620$-$00 and V404 Cygni effectively bookend the BHXB quiescent state (for currently known BHXBs). Yet, at a distance of $1.06\pm0.12$ kpc \citep{2010Cantrell}, compared to V404 Cygni's distance of $2.39\pm0.14$ kpc \citep{2009Miller_Jones}, A0620$-$00  still has a high enough flux density for a radio monitoring campaign. A0620$-$00 was discovered during an outburst  in 1975 \citep{1975Elvis}, and upon its return to quiescence was dynamically confirmed to contain a black hole ($6.61 \pm 0.25 M_{\odot}$) and a K-type star of mass $0.40 \pm 0.04 M_{\odot}$ \citep{mcclintock86, 2010Cantrell}, with an orbital period of  $7.75234 \pm 0.00010$ hours \citep{mcclintock86, gonzalez14}. Besides the 1975 outburst, only one other historical outburst is known from photographic plates in 1917 \citep{1976EachusWright}.  A radio outflow was first discovered from A0620$-$00 in quiescence by  \citet{2006GalloFenderMillerJones}, who reported a radio flux density of $51.1 \pm 6.9\,\mu$Jy at a central frequency of 8.5 GHz. The only other published radio detection of A0620$-$00 was by \citet{2018DincerBailyn}, who reported a radio flux density of $26 \pm 8\,\mu$Jy (8.5 GHz). The above indicates a factor of $\approx$2 variability.  However, this is based on only two observations taken eight years apart (observations in 2005 and 2013) with no radio detections between to bridge the observational gap.\footnote{To our knowledge, the only other published observation at GHz frequencies was in 2010 with the Australia Telescope Compact Array, which yielded a non-detection to a 5$\sigma_{\rm rms}$ limit of $<$80 $\mu$Jy \citep{2011_Froning}.} 
     It is thus unclear if the factor of $\approx$2 decrease in flux density represents a long-term dimming of the radio jet, or if the 2005 versus 2013 observations happened to catch A0620$-$00 while it exhibited different levels of (stochastic) variability. Subsequently, in 2016, the jet of A0620$-$00 was detected by \citet{2019GalloTeaguePlotkin} at millimetre wavelengths with the Atacama Large Millimeter/submillimeter Array (ALMA); they concluded that the quiescent jet displays variability in flux density and/or in spectral shape. 

    In this paper, we present  new radio observations of A0620$-$00 in quiescence taken with the VLA from 2017-2020. Our goal is to quantify the level of flux density variations from the quiescent jet to help inform future radio surveys of potential BHXB candidates, and to better understand the physical nature of the jet itself.
    This paper is arranged as follows: Section~\ref{sec:observations} describes the radio observations and data reduction, Section~\ref{sec:analysis_results} contains the results on the variability of the A0620$-$00 radio flux density,  results are discussed in Section~\ref{sec:discussion}, and a summary appears in Section~\ref{sec:summary}.

    \begin{table}
	\centering
	\caption{Summary of each radio observation included in this investigation. Column (1): the calendar date. Column (2): the Modified Julian Date. Column (3): the project code for the VLA. Column (4): the array configuration at the time of observation. Column (5): the total observing time on A0620$-$00 in minutes. Column (6): the flux density at 9.8 GHz in units of $\mu{\rm Jy}$. For non-detections, upper limits are presented at the $2\sigma_{\rm rms}$ level.}
	\label{tab:radio_obs}
	\begin{tabular}{lccccc} 
		\hline
		Date & MJD & Project & Config & $t_{\rm src}$ & $f_{9.8}$ \\
		 & & ID & & (min) & ($\mu \rm{Jy}$)\\
		\hline
		2017 Sept. 11 & 58007.5977 & 17B-233 & B & 20 & 28.2$\pm$5.6\\
		2017 Sept. 20 & 58016.5293 & 17B-233 & B & 20 & 21.7$\pm$6.3\\
		2017 Sept. 26 & 58022.4101 & 17B-233 & B & 20 & <12.7\\
		2017 Sept. 28 & 58024.6173 & 17B-233 & B & 21 & <14.4\\
		2017 Nov. 2 A & 58059.3162 & 17B-233 & B & 20 & <12.1\\
		2017 Nov. 2 B & 58059.4271 & 17B-233 & B & 20 & 15.8$\pm$7.0\\
		2017 Nov. 3 & 58060.3017 & 17B-233 & B & 20 & 17.0$\pm$6.7\\
		2017 Dec. 14 & 58101.2902 & 17B-233 & B & 38 & 11.7$\pm$3.5\\
		2017 Dec. 18 & 58105.2878 & 17B-233 & B & 92 & 12.7$\pm$3.0\\
		2017 Dec. 24 & 58111.2282 & 17B-233 & B & 92 & 15.4$\pm$2.3\\
		2018 Jan. 1 & 58119.2037 & 17B-233 & B & 92 & 13.3$\pm$2.4\\
		2018 Jan. 10 & 58128.2047 & 17B-233 & B & 92 & 23.2$\pm$3.7\\
		2018 Jan. 22 & 58140.1580 & 17B-233 & B & 114 & 14.1$\pm$2.3\\
		2019 Nov. 8 & 58795.4535 & 19B-001 & D & 87 & <10.0\\
		2019 Nov. 9 & 58796.4863 & 19B-001 & D & 85 & 29.8$\pm$4.1\\
		2019 Nov. 18 & 58805.3810 & 19B-001 & D & 85 & 16.5$\pm$3.7\\
		2019 Nov. 30 & 58817.4395 & 19B-001 & D & 87 & <10.4\\
		2019 Dec. 4 & 58821.5167 & 19B-001 & D & 87 & 11.1$\pm$4.1\\
		2019 Dec. 9 & 58826.4409 & 19B-001 & D & 87 & 21.7$\pm$3.8\\
		2019 Dec. 13 & 58830.2737 & 19B-001 & D & 85 & 12.0$\pm$3.9\\
		2019 Dec. 16 & 58833.3747 & 19B-001 & D & 87 & 7.8$\pm$3.4\\
		2019 Dec. 28 & 58845.2727 & 19B-001 & D & 85 & 10.5$\pm$3.1\\
		2020 Jan. 2 A & 58850.1150 & 19B-001 & D & 85 & 21.7$\pm$3.6\\
		2020 Jan. 2 B & 58850.2881 & 19B-001 & D & 87 & 21.5$\pm$3.9\\
		2020 Jan. 6 & 58854.1723 & 19B-001 & D & 85 & <7.4\\
		2020 Jan. 12 & 58860.1315 & 19B-001 & D & 85 & 20.8$\pm$3.3\\
		2020 Jan. 15 & 58863.2790 & 19B-001 & D & 87 & 9.0$\pm$3.9\\
		2020 Jan. 16 & 58864.1253 & 19B-001 & D & 85 & 15.6$\pm$4.2\\
		2020 Jan. 18 & 58866.0839 & 19B-001 & D & 85 & 10.3$\pm$4.3\\
		2020 Jan. 21 & 58869.2363 & 19B-001 & D & 87 & <8.5\\
		2020 Jan. 22 & 58870.0604 & 19B-001 & D & 85 & <9.1\\
		\hline
	\end{tabular}
\end{table}

\section{Radio Observations}
\label{sec:observations}

    We obtained a total of 31 observations of A0620$-$00 with the VLA over two separate observing programs, 17B-233 (13 observations executed from 2017 September$-$2018 January) and 19B-001 (18 observations from 2019 November$-$2020 January), which combined represent a total of 37 hours on source.  All of our data were taken in X-band (8-12 GHz), using 2$\times$2 GHz basebands centred at 9.0 and 10.65 GHz (these central frequencies were chosen to avoid radio frequency interference from satellites in the Clarke Belt).  During the 17B-233 observing campaign the VLA was in B configuration (maximum baseline 11.1 km), and during the 19B-001  campaign the VLA was in D configuration (maximum baseline 1.03 km).  A summary of our observations can be found in Table~\ref{tab:radio_obs}.
    
   The time on source for individual observations in the initial observing campaign in Table~\ref{tab:radio_obs} ranged between 20-114 minutes.  We began program 17B-233  with 40 minute observing blocks that yielded $\approx$20 minutes on source per observation.  This would have been a sufficient exposure time if A0620$-$00 displayed, on average, a radio flux density around 50 $\mu$Jy (i.e., similar to that observed by \citealt{2006GalloFenderMillerJones}). However, upon preliminary inspections of the data, A0620$-$00 was consistently more than 2--3 times fainter than expected, so we adapted to longer observations (at the expense of fewer epochs during 17B-233).  Ultimately, by the end of 17B-233 we settled on two hour observations that provided $\approx$85-90 minutes on source,  which were the observing times adopted for all epochs during the 19B-001 campaign. 
    
    During program 17B-233, 3C 48 was used as the primary calibrator to set the flux density scale, to perform delay calibrations, and to find complex bandpass solutions for 11/13 observations (3C 286 was  the primary calibrator for the other two observations, with the choice of calibrator dictated by the local sidereal time at the beginning of each observation). Note, 3C 48 experienced a flare starting around January 2018, which may affect the flux density scale at the 5-10\% level for the final $\approx$3 observations during the 17B-233 campaign.\footnote{\url{https://science.nrao.edu/facilities/vla/docs/manuals/oss/performance/fdscale}}
    For the 19B-001 campaign, 3C 147 was used as the primary calibrator for all 18 observations.  Across both campaigns,  J0641$-$0320 was used as the secondary calibrator, where we cycled to J0641$-$0320 every $\approx$10-15 minutes to solve for temporal changes in the complex gain solutions.

  Data were calibrated and imaged according to standard procedures within the Common Astronomy Software Application v5.6.2-3 \citep[{\tt CASA};][]{2007McMullin}. 
    Initial calibrations were performed using the VLA pipeline (v5.6.2-2), followed by manual inspection of the calibrated measurement sets.  In most observations, data between 11-12 GHz were heavily flagged due to interference from the Clarke Belt.  If flagging was extensive, then the data were re-run through the pipeline for re-calibration.  Data were then imaged using {\tt tclean}, adopting two Taylor terms ({\tt nterms=2}) to account for spectral dependences across the large fractional bandwidth. During the 17B-233 campaign, natural weighting was used to maximise sensitivity in the more extended B configuration (typical synthesised beam of $\theta_{\rm HPBW} \approx 0.6$ arcsec), and for the 19B-001 campaign Briggs weighting with {\tt robust}=0.5 was used to balance sensitivity with mitigating bright side lobe artifacts in the more compact D configuration (typical synthesised beam of $\theta_{\rm HPBW} \approx 7.2$ arcsec).

    The flux density of A0620$-$00 was measured by fitting a point source in the image plane at the known location of A0620$-$00 (but allowing the position of the centre of the point source to vary). 
    For the 17B-233 program, we achieved  average rms noise levels of $\approx$$6.0\,\mu {\rm Jy\,bm^{-1}}$ for observations with 20 minutes on source, $\approx$$3.5\,\mu {\rm Jy\,bm^{-1}}$ for 40 minutes on source, $\approx$$3.0\,\mu {\rm Jy\,bm^{-1}}$ for 90 minutes on source, and $2.3\,\mu {\rm Jy\,bm^{-1}}$ for the lone observation with 114 minutes on source. For the 19B-001 program, we achieved an average rms of $\approx$$4.0\,\mu {\rm Jy\,bm^{-1}}$ for 90 minutes on source (the slightly worse $\sigma_{\rm rms}$ in 19B-001 compared to epochs  during the 17B-233 program with similar exposure times is due to the more compact D configuration, and the use of Briggs versus natural weighting). Table~\ref{tab:radio_obs} lists  measured flux densities for each observation.  Since the position of A0620$-$00 is well-known and radio emission has been detected previously from  A0620$-$00 in quiescence \citep{2006GalloFenderMillerJones}, we adopt a detection threshold of $2 \sigma_{\rm rms}$ throughout this paper. We note that this is lower than the more traditionally adopted 3$\sigma_{\rm rms}$  threshold for detection experiments.  However, if the noise characteristics in the radio maps are dominated by statistical fluctuations, then the 2$\sigma_{\rm rms}$ threshold risks only a 5\% chance of misidentifying a nearby noise spike as a detection.   Only six of our detections fall in the $2-3 \sigma_{\rm rms}$ range, implying that adopting a 2$\sigma_{\rm rms}$ threshold instead of 3$\sigma_{\rm rms}$ may introduce 0.3 spurious detections (i.e., at most one false detection), which would not influence our final results.  Furthermore, we find that all six of the $2-3 \sigma_{\rm rms}$ detections fall within 1--2 synthesized beams of the known location of A0620$-$00, providing further confidence that these six detections are robust.
    
 Finally, we attempted to measure in-band spectral indices, $\alpha$ (where $f_\nu \propto \nu^\alpha$), for our brightest detections by splitting  observations into subbands centred at 8.7 and 10.1 GHz.  However, given the relatively narrow lever arm in frequency and low signal-to-noise, spectral indices have large uncertainties ($\sigma_\alpha > \pm 1$) and are not useful.  We therefore do not report spectral information in this paper.

    \begin{figure}
        \centering
        \includegraphics[width=1.0\columnwidth]{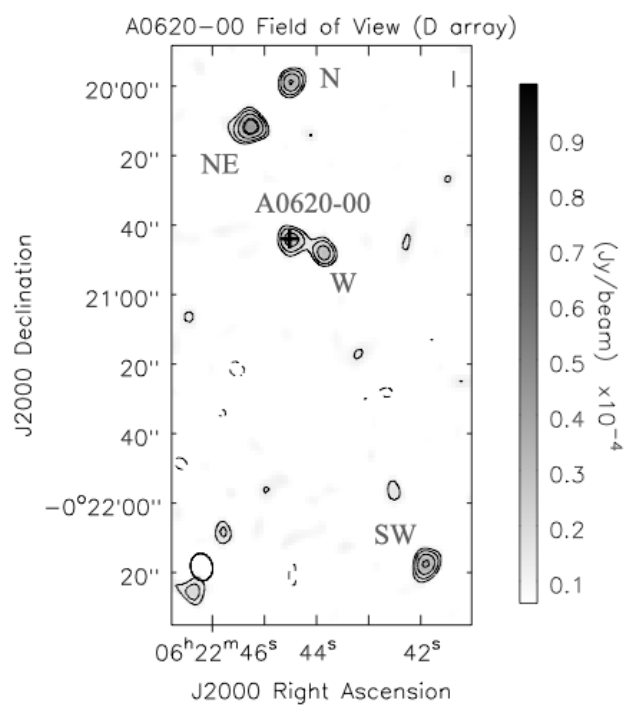}
        \caption{9.8 GHz radio image overlaid with radio contours of the field of view surrounding A0620$-$00 as observed by the VLA on 2019 November 09. Contours are drawn at $\pm (\sqrt{3}^n)$ times the rms noise level of 4.0 $\mu {\rm Jy\,bm}^{-1}$, where $n = 2,3,4,5,..$. A0620$-$00 is clearly detected and distinguished with cross hairs. The N and NE sources mentioned by \citet{2006GalloFenderMillerJones} are also clearly detected and labeled. Additionally, W and SW sources are detected and labeled in this image as they are used as check sources during our analysis. 
        }
        \label{fig:fov}
    \end{figure}

      \begin{figure*}
        \centering
        \includegraphics[width=1.0\linewidth]{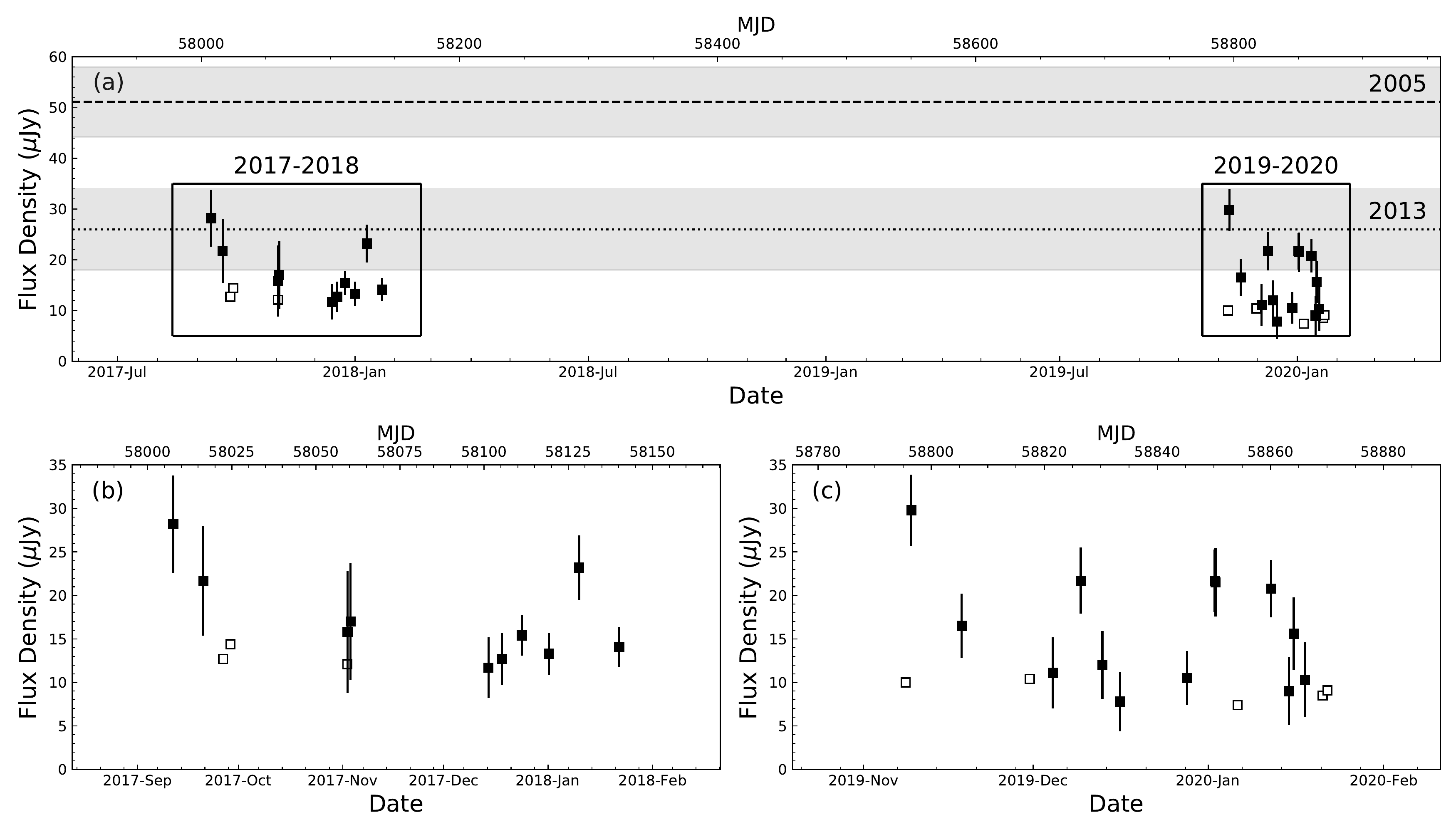}
        \caption{ (a) Radio light curve of A0620$-$00 from  2017 September through  2020 January, at 9.8 GHz. Filled symbols represent detections, and open symbols represent 2$\sigma_{\rm rms}$ upper limits on non-detections. 
        The horizontal dashed line marks the flux density observed in 2005 at 8.5 GHz \citep{2006GalloFenderMillerJones}, and the horizontal dotted line marks the flux density observed in 2013 at 8.5 GHz \citep{2018DincerBailyn}; the grey shaded regions illustrate the 1$
        \sigma_{\rm rms}$ uncertainties of those measurements. (b) A zoom-in of the flux density variations during the 17B-233 observing campaign, spanning 2017 September - 2018 January. The data with larger error bars correspond to shorter times on source, which ranged from 20--114 minutes (see Table~\ref{tab:radio_obs}).  (c) Data from the 19B-001 observing campaign, spanning 2019 November - 2020 January. All of these observations had approximately the same time on source ($\approx$90 minutes).}
        \label{fig:all_time_lightcurve}
    \end{figure*}

    \section{Results}
    \label{sec:analysis_results}

    Figure~\ref{fig:fov} shows an example radio image of one of our strongest radio detections in D configuration ($29.8\pm4.1$ $\mu$Jy on 2019 November 9), to illustrate the field of view of A0620$-$00, and also to note the locations of four nearby field sources  that we utilised as `check sources.'   
    A0620$-$00 did not vary in tandem with any of our check sources, indicating that observed variations of A0620$-$00  reflect real changes in the flux density.  In Appendix \ref{appendix} we show images for all 31 individual observations. Extended emission is never observed near A0620$-$00, justifying our choice to force a point source while performing flux density measurements in {\tt imfit}. Applying our 2$\sigma_{\rm rms}$ detection threshold, eight out of the thirty-one observations  are considered non-detections. We stacked the three non-detections from the 17B-233 campaign and obtained a $2\sigma_{\rm rms}$ upper limit of 6.6 $\mu$Jy. Similarly, we stacked the five non-detections from the 19B-001 campaign and obtained an upper limit of 3.7 $\mu$Jy.

\begin{figure}
            \centering
            \includegraphics[width=1.0\columnwidth]{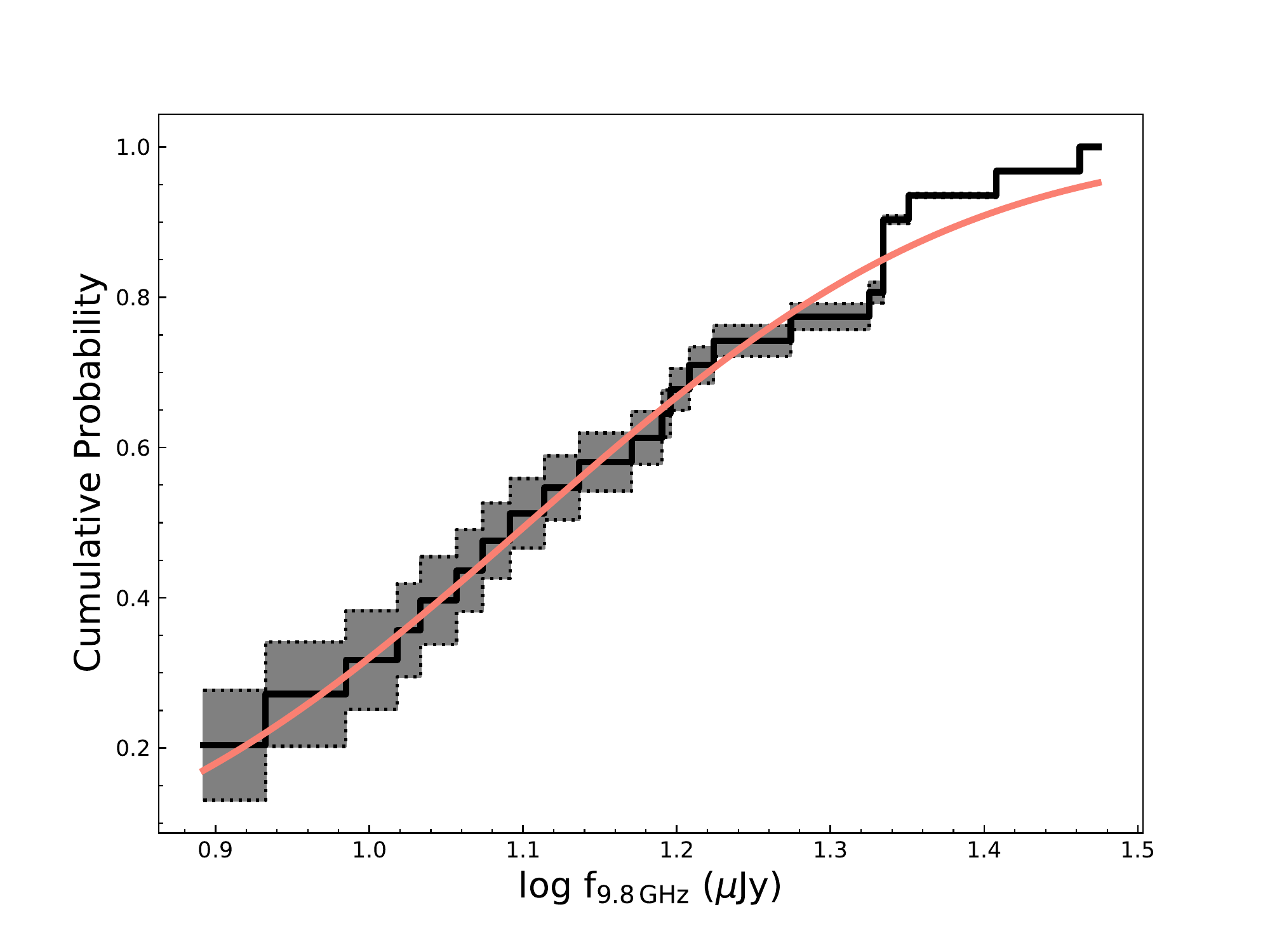}
            \caption{Cumulative distribution function (solid black histogram) of the logarithm of flux densities (9.8 GHz) for our 31 observations, plotted as $1-S(\log f_{\rm 9.8}$), where $S(\log f_{\rm 9.8})$ is the survival function from the Kaplan-Meier estimator.  The gray shaded region represents the $\pm 1\sigma$ uncertainty about $1-S(\log f_{\rm 9.8})$.  The flux density is consistent with a lognormal distribution with mean $\left<\log\left(f_{\nu}/{\mu{\rm Jy}}\right)\right> = 1.10$ and standard deviation $\pm$0.22 dex, depicted by the red line.}
            \label{fig:kap_mei}
        \end{figure}

    The 9.8 GHz light curve spanning both observing campaigns is shown in Figure~\ref{fig:all_time_lightcurve}. To quantify the distribution of flux densities observed across our entire campaign, including information from non-detections, we performed a survival analysis. We used the survival  package {\tt survfit} in {\tt R} to calculate the Kaplan-Meier product-limit estimator of the logarithms of the flux densities \citep[see, e.g.,][]{1985KaplanMeier}, which calculates the   survival function, $S \left({\rm log}f_{\rm 9.8}\right)$.  We display this result as $1-S \left({\rm log}f_{\rm 9.8}\right)$ in Figure~\ref{fig:kap_mei}, which represents a quantity similar to a cumulative distribution function.
    The median value of the survival function is $\log \left(f_\nu/\mu{\rm Jy}\right)=1.10$, corresponding to $\log (L_{\rm R}/{\rm erg\,s}^{-1})=26.2$. The 16-84 percentiles of the survival function provide an estimate of $\pm0.22$ dex for the $\pm 1\sigma$ confidence interval.  We  compare the observed survival function to a lognormal distribution through a Peto \& Peto test,\footnote{Specifically, we use the Peto \& Peto modification of the Gehan–Wilcoxon test implemented by {\tt cendiff} in the {\tt R} package {\tt Nondetects and Data Analysis for Environmental Data (NADA).}} 
    and we find the probability for the null assumption (i.e., that the observed flux densities follow a lognormal distrubtion) to be $p=0.83$.  Thus, the  distribution of radio flux densities is consistent with a lognormal distribution with mean $\left< \log \left(f_\nu/\mu{\rm Jy}\right)\right>=1.10$ and standard deviation of $\pm$0.22 dex (note, the $\pm$0.22 dex uncertainty quantifies the level of flux variability and is not meant to represent the error on the mean).  
    The fractional variability for the 23 detections is $F_{var} = 26 \pm 8\%$ \citep{2003Vaughan}, which represents a lower limit on the level of variability.

    To further quantify variability we calculated the first-order structure function V$\left(\tau\right)$, which is one way to quantify variations in irregularly sampled data. We define the structure functions as:
        
        \begin{equation}
            V\left(\tau\right) = \left< \left[f\left(t + \tau\right) - f\left(t\right)\right]^2 \right>
            \label{eq:sf}
        \end{equation}
        where $f(t)$ is the flux density at time $t$, and $\tau$ is the time difference between two observations. 
        Due to the different exposure times and resulting different $\sigma_{\rm rms}$ noise levels, we were unable to calculate a useful structure function for the 2017-2018 data. In a similar vein, we did not calculate the structure function for the data spanning both observing campaigns due to the non-uniform error bars when comparing the two campaigns. Since non-detections cannot be incorporated into the structure function, the following analysis included only the 13 detections from the 19B-001 observing campaign.  We stress that this is a very small sample size for attempting a structure function analysis, but in this case it is sufficient to gain some preliminary insight into the variability characteristics of A0620$-$00.

         \begin{figure}
            \centering
            \includegraphics[width=1.0\columnwidth]{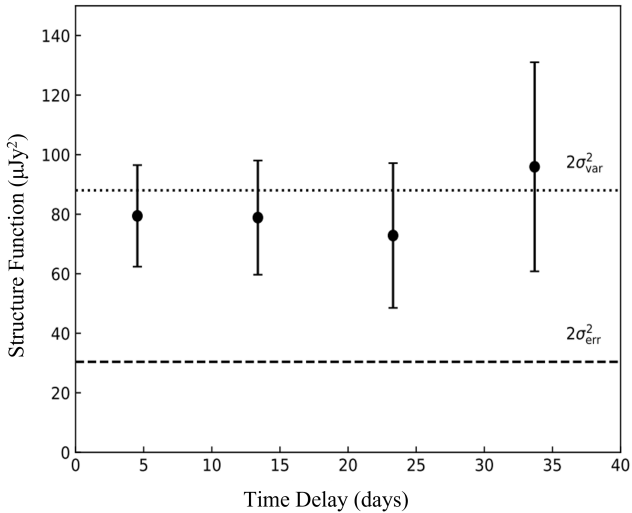}
            \caption{The first order structure function, including only the data from 2019-2020  (13 measurements, omitting non-detections). The structure function is binned to include 15 pairs of data points per time delay. The dashed horizontal line represents twice the average measurement error squared $(2\sigma^2_{\rm err})$ and the dotted horizontal line represents twice the variance $(2\sigma^2_{\rm var})$ of the 13 data points. Note that the dotted line is not a fit to the structure function. Flux density variations are in excess of statistical fluctuations from measurement errors. The flat slope represents either flicker noise or white noise on timescales longer than $\approx$5 days.}
            \label{fig:structure_funct}
        \end{figure}
        
        Although the variations follow a lognormal distribution, we calculated the structure function in linear space to maximise the dynamic range.  For every pair of detections $(t_i, t_j)$ during  the 19B-001  campaign, we defined $\tau_{ij} = t_j - t_i$ and  $V_{ij} \left(\tau_{ij} \right) = \left[f\left(t_j\right) - f\left(t_i\right)\right]^2$. The set of $V_{ij}$ values was then binned by time delay so that each bin had 15 data points and $V\left(\tau \right)$ was calculated as the average across the data contained within each bin, where $\tau$ is the midpoint of the time delays  within each bin. Figure~\ref{fig:structure_funct} shows the resultant structure function examining variability over timescales $\approx$ 5-35 days (the 19B-001 campaign lasted $\approx$75 days, and one can typically only quantify the structure function to within $\approx$half of the maximum time baseline).  We fitted a line to the structure function and found it is consistent with a flat slope $\beta=0.055 \pm 0.190$ that plateaus to a value consistent with 2$\sigma^2_{\rm var} = 88.0\, \mu {\rm Jy}^{2}$, where  $\sigma^2_{\rm var}$ is the variance of the measured flux densities. 

\section{Discussion}
    \label{sec:discussion}

    We present light curves of A0620$-$00 in quiescence from 31 VLA observations spanning 2017 September - 2020 January. In the following, we interpret all radio detections as emission from a compact jet. As discussed in \citet{2006GalloFenderMillerJones}, the radio flux density of A0620$-$00 is too large to be attributed to other sources of emission (e.g., the underlying accretion flow, coronal emission from the companion star, etc.).  Even though A0620$-$00 is 2--3 fainter during our observing campaigns compared to the $\approx$50 $\mu$Jy detection in \citet{2006GalloFenderMillerJones}, their conclusions still hold.  For example, consider our faintest radio detection, which is 7.8 $\mu$Jy on 2019 December 16.  If this emission were to be produced by an advection dominated accretion flow \citep[ADAF;][]{ichimaru77, narayan95, yuan14}, the physical size of the emission region would be on the order of 300 gravitational radii \citep[e.g.,][]{Veledina2013}.  In that case, the brightness temperature would be  $\approx$10$^{14}$ K, which is significantly above the limit from the inverse Compton catastrophe. It is therefore not physically possible for an ADAF to produce enough radio emission.
    
    Below, in Section \ref{sec_short_term} we interpret the short timescale variability, i.e. days to years, over both observing campaigns. In Section \ref{sec_long_term} we discuss our campaigns in context with previous observations over decade timescales.
    
    \subsection{Short-term behaviour of the radio jet}
    \label{sec_short_term}
    
    Over both campaigns, we find that flux density variations by a factor of 2-4 are common on timescales of days through years. A pair of observations from 2019 November 8--9 represents one the most extreme variations observed during our campaign, where A0620$-$00 went undetected on November 8 ($<$10.0 $\mu$Jy), followed by the brightest flux density observed during our campaign 25 hours later ($29.8\pm4.1$ $\mu$Jy).
    
    From the Kaplan-Meier estimator (Figure~\ref{fig:kap_mei}),  typical $\pm$1$\sigma$ variations of A0620$-$00 at 9.8 GHz are $\pm$0.22 dex.  The structure function of A0620$-$00 is flat and plateaus to 2$\sigma_{\rm var}^2$, which (a) confirms that the variations are intrinsic (i.e., they are in excess of expectations from statistical noise related to measurement error) and (b) the long-term fluctuations on 5-35 day timescales are consistent with either flicker or white noise (see Section 3.1 of \citealt{2019PlotkinMillerJones}, and references therein, for interpretations of flat structure functions).  Note, these timescales are shorter than the viscous time of the outer accretion disk, thereby implying that the radio jet variability is not mimicking changes in the mass transfer rate from the donor star to the outer disk. Qualitatively, these results are similar to those obtained by \citet{2019PlotkinMillerJones} for the radio variability of V404 Cygni in quiescence (note, their structure function spanned $\approx$10-4000 days instead of $\approx$5-35 days for A0620$-$00).  In the case of V404 Cygni, a small number of flares were also observed on minute-to-hour timescales, such that \citet{2019PlotkinMillerJones} concluded that the variability of V404 Cygni is  consistent with a damped random walk, likely related to shocks propagating through a jet \citep[e.g.,][]{malzac14}.   Even though A0620$-$00 is too faint to  resolve individual flares on timescales less than 1-2 hours, a similar explanation is likely.  In particular, \citet{2018DincerBailyn} saw potential variability on a timescale of $\approx$2.5 hours, where the 22 GHz flux density of A0620$-$00 increased from $\approx$$50\pm10$  to $\approx$$80\pm10$ $\mu$Jy.\footnote{Our campaign was generally insensitive to intraday variability, except for a  pair of observations taken on 2017 November 2 separated by $\approx$2.6 hours, and a pair on 2020 January 2 separated by $\approx$4 hours.  Only the 2017 November 2 observations may hint at short-term variability, where we obtained a non-detection ($<$12.1 $\mu$Jy) followed by a detection ($15.8\pm7.0$ $\mu$Jy), but we do not consider this to be statistically significant.} 
    If variations from the radio jet of A0620$-$00 indeed follow a damped random walk, then the structure function places a limit of less than 5 days on the damping timescale.
    
\begin{figure}
    \centering
    \includegraphics[width=1.0\columnwidth]{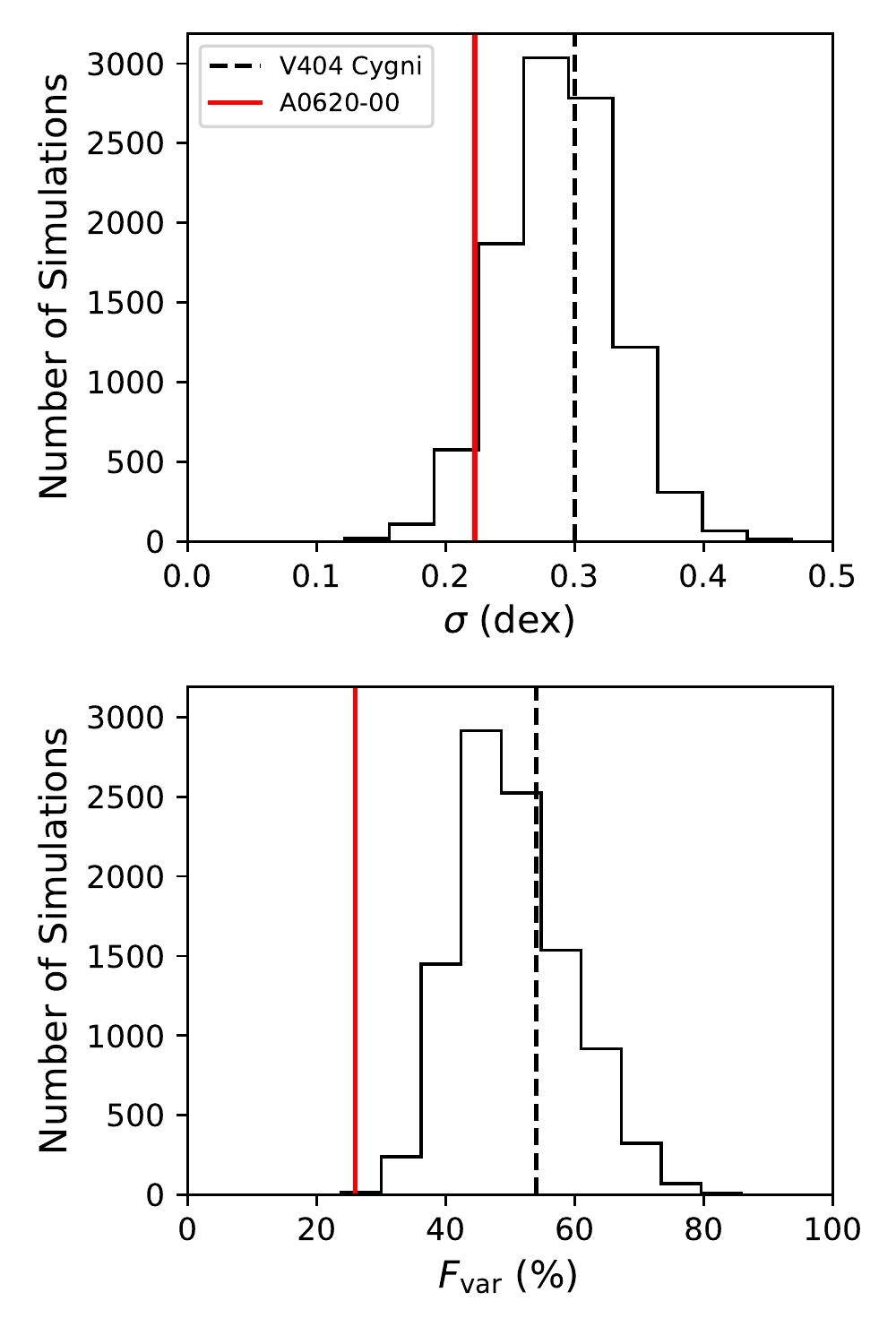}
    \caption{Black histograms show results from 10000 Monte Carlo simulations randomly drawing 31 data points from the 8.4 GHz radio light curve of V404 Cygni from \citet{2019PlotkinMillerJones}.  \textit{Top panel:} the standard deviation of each realisation of the V404 Cygni light curve, parameterised by the 16--84 percentile range of the survival function from the Kaplan-Meier distribution.  The vertical black dashed line shows the value calculated by \citet{2019PlotkinMillerJones} from the full light curve, and the vertical solid red line shows the value calculated for A0620$-$00 in this paper.  \textit{Bottom panel:} the distributions of $F_{\rm var}$ values calculated from the 10000 Monte Carlo simulations.  The black dashed and solid red vertical lines have the same meanings as in the top panel.  These simulations suggest that the lower level of variability observed from A0620$-$00 compared to V404 Cygni is unlikely an artifact of the poorer sampling of the A0620$-$00 light curve, especially for $F_{\rm var}$.}
    \label{fig:mcs}
    \end{figure}

   Despite the above qualitative similarities, there may be  quantitative differences in that A0620$-$00 appears to display a lower degree of fractional variability (at 9.8 GHz) compared to V404 Cygni (at 8.4 GHz).  The standard deviation of the lognormal flux density distribution of A0620$-$00 is $\pm$0.22 dex compared to $\pm$0.30 dex for V404 Cygni, and, more significantly, the fractional variability (in linear space) of A0620$-$00 is $F_{\rm var} = 26\pm8$\% compared to $F_{\rm var} = 54 \pm 6$\% for V404 Cygni \citep{2019PlotkinMillerJones}.  We note that A0620$-$00 has $\approx$3 times fewer data points in its quiescent radio light curve compared to the 8.4 GHz light curve of V404 Cygni.  To investigate whether the smaller standard deviation and $F_{\rm var}$ from A0620$-$00 is  an artifact of poorer sampling of its flux density distribution, we run Monte Carlo simulations by randomly selecting 31 data points from the 8.4 GHz light curve of V404 Cygni (89 data points)  published by \citet{2019PlotkinMillerJones}.\footnote{Note, there is insufficient coverage of V404 Cygni for us to also attempt to replicate the observing cadence of A0620$-$00 over a 3--4 month period.  Thus we do not attempt to simulate the effect that fewer data points will have on the structure function.}  
   For these 31 data points, we calculate the Kaplan-Meier estimator (incorporating non-detections) and $F_{\rm var}$ (including only detections), and we repeat 10000 times.  Results are shown in Figure~\ref{fig:mcs}.

   The above realisations of the data with only 31 data points  recover, on average, similar standard deviations of the lognormal flux density distribution ($0.29\pm0.04$ dex) compared to the full V404 Cygni light curve with 89 data points ($\pm$0.30 dex).  Similar values for $F_{\rm var}$ are also found, on average, from  our simulations ($50\pm9$\%) compared to the full light curve (54\%).  Thus, sampling a light curve with only 31 data points does not appear to significantly bias our results, within the uncertainties.  Although, a small caveat is that the distribution of $F_{\rm var}$ values recovered by our simulations are asymmetric (see Figure~\ref{fig:mcs}).  The mode from our simulations is 46\%,  which is slightly smaller than the value calculated from the full light curve (54\%), albeit not at a level we consider highly significant (see further discussion below). 
   
    If A0620$-$00 were to have the same level of intrinsic variability as V404 Cygni, then the above simulations also allow us to quantify the probability that our campaign on A0620$-$00 happened to underestimate its intrinsic variability by random chance.  We find that 6.0\% of our simulations recover a standard deviation as small (or smaller) than the value observed from A0620$-$00 ($\pm$0.22 dex), and 0.03\% of our simulations recover an $F_{\rm var}$ value as small as observed from A0620$-$00 (26\%). Thus, if A0620$-$00 has similar levels of intrinsic radio variability as V404 Cygni, it is plausible that the poorer sampling of the A0620$-$00 light curve could yield the smaller observed standard deviation; however, the poorer sampling is very unlikely the primary driver for the lower observed value of $F_{\rm var}$.  Note, both the A0620$-$00 and V404 Cygni light curves have a similar fraction of non-detections (8/31=26\% for A0620$-$00 versus 25/89=28\% for V404 Cygni), so the above results are unlikely suffering from systematics related to the survival analysis and/or only including detections in the $F_{\rm var}$ calculation.  Thus, A0620$-$00 appears to indeed be (slightly) less variable than V404 Cygni.
     This result could imply that lower luminosity jets are less variable. Confirmation will require performing similar campaigns on other quiescent BHXBs (which will require  more sensitive radio facilities, and/or discovering new BHXBs at close enough distances to detect their quiescent radio jets; see, e.g., \citealt{rodriguez20}).  Confirmation of this  result would have favourable implications for variability inducing less severe systematics (at low luminosities) when searching for  new BHXB candidates via their radio jets.
    
    \subsection{Long-term behaviour of the radio jet}
    \label{sec_long_term}
    
    We conclude the discussion by using our variability constraints to attempt to better understand the long-term behaviour of A0620$-$00's radio jet.  Previous detections of the radio outflow (at GHz frequencies) include the $51.1 \pm 6.9\,\mu {\rm Jy\,bm^{-1}}$ detection in 2005 at 8.5 GHz by \citet{2006GalloFenderMillerJones}, 
    and the $26\pm8$ $\mu$Jy detection in 2013 at 8.5 GHz, based on the broadband spectrum observed by \citet{2018DincerBailyn}.
    The average 9.8 GHz flux density from our 17B-233 program  (2017 September -- 2018 January) is $17.3 \pm 1.5\,\mu {\rm Jy\,bm^{-1}}$, and the average flux density from our 19B-001 program (2019 November -- 2020 January) is $16.0 \pm 1.1\,\mu {\rm Jy\,bm^{-1}}$ (note, the uncertainties reported for these average flux densities are the error on the mean).  
    
    It is striking how much brighter the radio jet was during its discovery in 2005, and that over $\approx$30 subsequent observations (covering nearly 15 years) the jet has never been observed to be so bright again. One interpretation is a gradual, long-term dimming of the radio jet (the differences in observing frequencies above would result in additional $\lesssim$10\% systematic uncertainties, adopting the inverted spectral index $\alpha = 0.74 \pm 0.19$ measured by \citealt{2018DincerBailyn}). Alternatively, stochastic variability cannot be excluded, which would imply that the discovery in 2005 fortuitously caught A0620$-$00 during a relatively bright flare. The 2005 observation, $\log \left(f_{\rm 9.8}/\mu{\rm Jy}\right) = 1.71$, represents a 3$\sigma$ deviation from the mean of our Kaplan-Meier distribution, thus yielding a 0.15\% chance of finding the source at least this bright by random chance (assuming one can safely extrapolate our observed flux density distribution to higher flux densities). This is a small but  non-negligible probability. For the 2013 observations, interpolating the flux density reported by \citet{2018DincerBailyn} from 8.5  to 9.8 GHz, that observation would have $\log \left(f_{\rm 9.8}/\mu{\rm Jy}\right) = 1.46$.  Our Kaplan-Meier distribution suggests a 2.8\% chance of observing A0620$-$00 at $\log \left(f_{\rm 9.8}/\mu{\rm Jy}\right) > 1.46$.  
    
    Our constraints on the level of variability at 9.8 GHz are also helpful for interpreting broadband spectra of quiescent jets.  For example, \citet{2019GalloTeaguePlotkin} detected an apparently steep sub-millimetre spectrum from A0620$-$00 using the ALMA, based on detections at 98 GHz ($44\pm 7$ $\mu$Jy) and at 233 GHz ($20 \pm 8$ $\mu$Jy).  As pointed out by \citet{2019GalloTeaguePlotkin}, the observations at these two frequencies were performed 40 days apart, such that  a factor of $\approx$2 variability would be sufficient to influence the observed sub-millimetre spectrum.  Given that such variability is common from A0620$-$00 at GHz frequencies, our campaign supports the conclusion of \citet{2019GalloTeaguePlotkin} that jet variability is one plausible explanation, especially when considering that compact synchrotron jets are expected to be more variable at higher frequencies \citep{tetarenko19, tetarenko21}.

\section{Summary}
    \label{sec:summary}

    We presented a  monitoring campaign on the radio jet launched by the quiescent BHXB A0620$-$00,  spanning 2017 September -- 2018 January and 2019 November -- 2020 January, with a total of 31 observations representing 37 hours on source. These observations comprise the most intense radio monitoring campaign yet on A0620$-$00, which we use to provide statistical constraints on the radio variability of one of the least luminous BHXB jets.  We find  flux densities that follow a lognormal distribution, with a mean $\left<\log \left(f_\nu/\mu{\rm Jy}\right)\right>=1.10$ and a standard deviation of $\pm$0.22 dex at 9.8 GHz (the average flux density corresponds to a radio luminosity at 9.8 GHz of $L_{\rm R}\approx 2 \times10^{26}$ erg s$^{-1}$).   Factor of 2--4 flux density variations are common on timescales ranging from days to years, and we quantify the fractional variability to be at least $F_{\rm var}=26 \pm8$\%.  Over the data collected between 2019--2020, A0620$-$00 displays a flat structure function, indicating that the observed flux density variations  are consistent with either flicker noise or white noise on timescales 5--35 days. Intriguingly, none of our observations found A0620$-$00 to be as bright as during its discovery observation in 2005, implying either a long-term dimming of the radio jet, or the 2005 observations happened to catch A06260$-$00 during a relatively bright flare. The variability characteristics of A0620$-$00 are similar to V404 Cygni, the only other BHXB for which radio variability has been similarly quantified in quiescence, except that A0620$-$00 may be slightly less variable (e.g., V404 Cygni follows a lognormal flux density distribution at 8.4 GHz with a standard deviation of $\approx \pm 0.3$ dex and  $F_{\rm var} = 54\pm6$\%).  This lower level of variability from A0620$-$00 may indicate that lower luminosity quiescent jets could be less variable at GHz frequencies.  
    
    Understanding the level of jet variability is important for being able to interpret broadband spectra built with non-simultaneous multiwavelengh data (e.g., \citealt{gallo07, 2019GalloTeaguePlotkin}), and it also has implications for using radio surveys, in conjunction with non-simultaneous multiwavelength information, to uncover new populations of quiescent BHXBs \citep[e.g.,][]{2005Maccarone}.  Further monitoring of A0620$-$00 will be useful to better sample  its flux density distribution and improve our statistical calculations, and it would be worthwhile to repeat such a study at other frequencies. To improve our understanding of quiescent BHXB jet variability, however, will ultimately require discovering more nearby systems, or else await new generations of radio telescopes such as a next generation VLA (ngVLA) or the Square Kilometre Array (SKA-MID).  A facility like an ngVLA in particular, which would have an order of magnitude improvement in sensitivity compared to the VLA, would allow us to significantly expand the number of feasible targets, while also opening up searches for flaring activity (on minute-to-hour time scales) from jets as faint as A0620$-$00.

\section*{Acknowledgements}

We thank the anonymous referee for constructive comments that improved the manuscript. The National Radio Astronomy Observatory is a facility of the National Science Foundation operated under cooperative agreement by Associated Universities, Inc.  This research made use of Astropy,\footnote{http://www.astropy.org} a community-developed core Python package for Astronomy \citep{astropy-collaboration13, astropy-collaboration18}. J.C.A.M.-J. was supported by an Australian Research Council Future Fellowship (FT140101082). J.S.~acknowledges support from the Packard Foundation.

\section*{Data Availability}

The data that supported the findings of this study are available at the National Radio Astronomy Observatory archive at \url{data.nrao.edu} under the project codes: 17B-233 and 19B-001. 
 



\bibliographystyle{mnras}
\bibliography{ref} 



\appendix

    \section{Images of Individual Observations}
        \label{appendix}

\begin{figure*}
       \centering
        \includegraphics[width=0.95\linewidth]{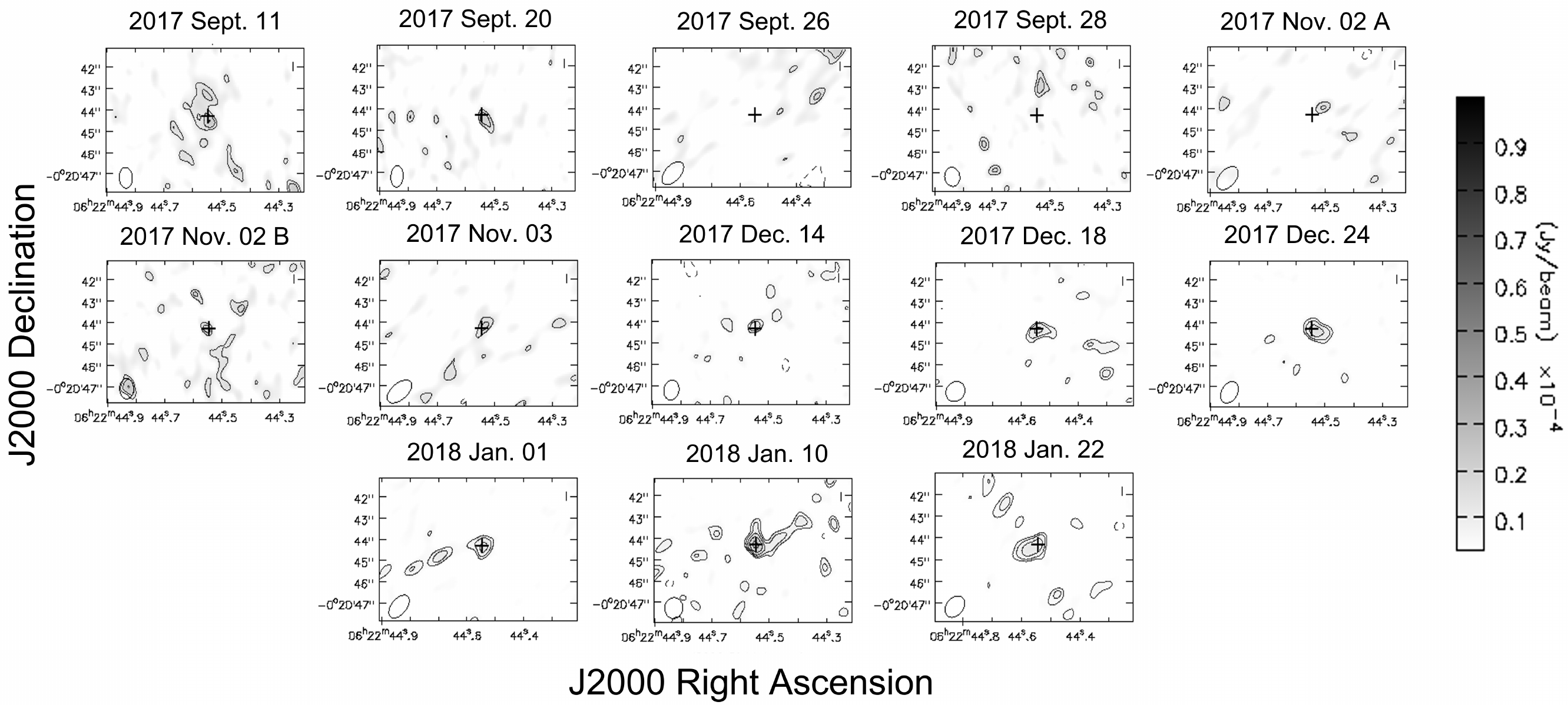}
       \caption{Contour images summarising the detections and non-detections of A0620$-$00 from the 17B-233 observing campaign. The coordinates of A0620$-$00 are marked in each panel with cross hairs. Since the images from 2017-2018 had different exposure times, the contours for each panel correspond to the average $\sigma_{\rm rms}$ with respect to the exposure time. For 20 minutes time on source (t.o.s.) (2017 September 11 - 2017 November 3), contours are at $\pm (\sqrt{2}^n)$ times the $\sigma_{\rm rms}$ of $6.0\,\mu {\rm Jy\,bm}^{-1}$, with $n = 2,3,4,5,...$. For 40 minutes t.o.s (2017 December 14), contours are similarly scaled to $\sigma_{\rm rms}= 3.5\,\mu {\rm Jy\,bm}^{-1}$. For 90 minutes t.o.s. (2017 December 18 - 2018 January 10), contours are  scaled to $\sigma_{\rm rms}= 3.0\,\mu {\rm Jy\,bm}^{-1}$. Finally, for 115 minutes t.o.s. (2018 January 22), contours are  scaled to $\sigma_{\rm rms}= 2.3\,\mu {\rm Jy\,bm}^{-1}$. The non-detections for this campaign are 2017 September 26, 2017 September 28, and 2017 November 02 A.}
        \label{fig:detect17}

    

       \centering
        \includegraphics[width=0.95\linewidth]{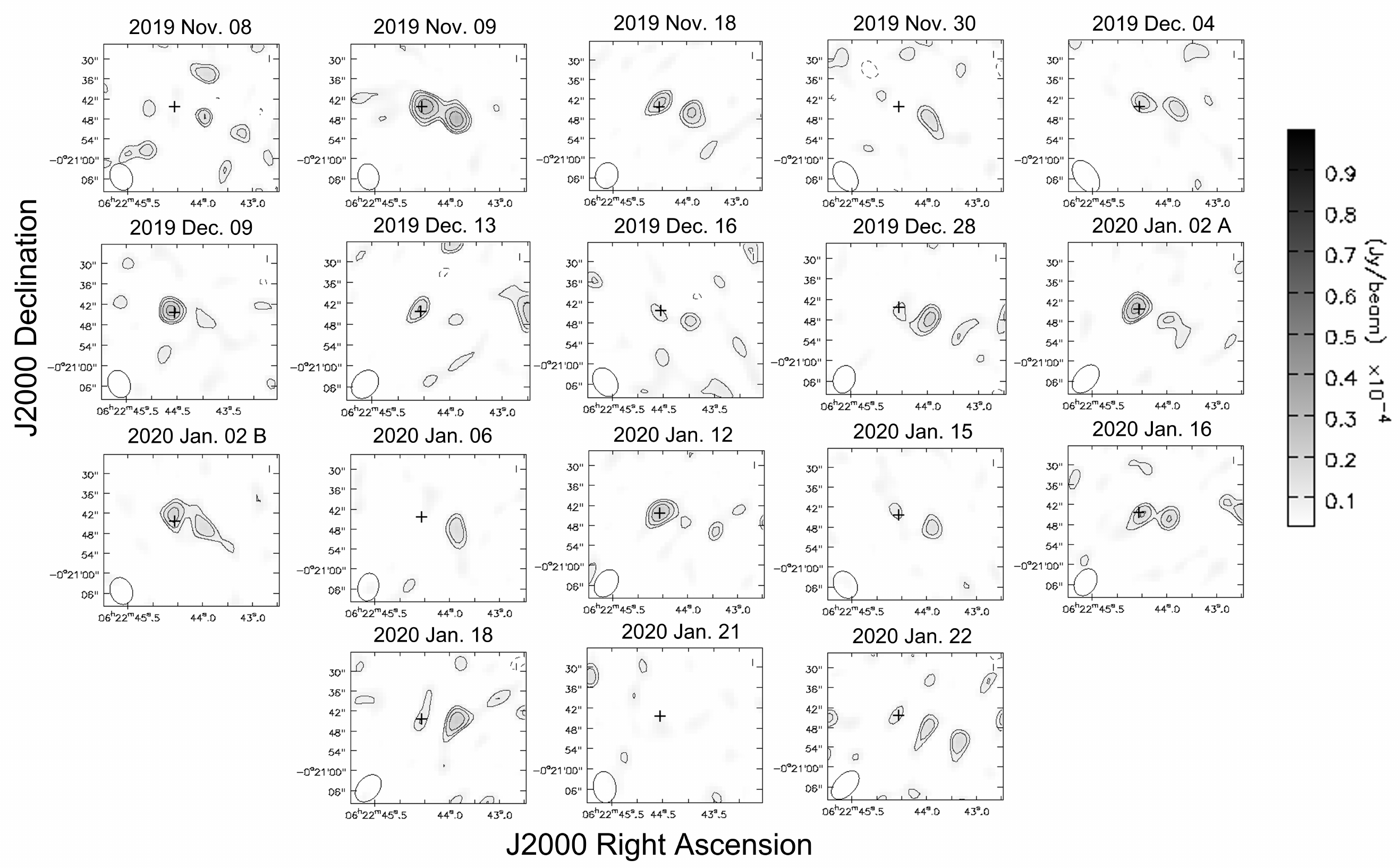}
       \caption{Contour images summarising the detections and non-detections of A0620$-$00 from the 19B-001 observing campaign. The coordinates of A0620$-$00 are marked in each panel with cross hairs. Since the images from 2019-2020 all the had the same time on source of 90 minutes, the contours are at $\pm (\sqrt{2}^n)$ times the the $\sigma_{\rm rms}$ of $4.0\,\mu {\rm Jy\, beam}^{-1}$, with $n=2,3,4,5,...$.The non-detections for this campaign are 2019 November 08, 2019 November 30, 2020 January 06, 2020 January 21, and 2020 January 22. Note that in the panel above, 2020 January 22 appears to be a detection, however $\sigma_{\rm rms}$ for this image was higher than average, making the appearance of a detection in the image when the flux density was actually below the $2\sigma$ limit on this particular date.}
        \label{fig:detect19}
    \end{figure*}

    Below we show images of each observation in our 17B-233 (Figure \ref{fig:detect17}) and 19B-001 campaigns (Figure \ref{fig:detect19}). 
   

\bsp	
\label{lastpage}
\end{document}